# Effects of Tonal Coarticulation and Prosodic Positions on Tonal Contours of Low Rising Tones: In the Case of Xiamen Dialect

*Yiying Hu[1], Hui Feng[1*], Qinghua Zhao[1], Aijun Li[2]*

[1]Tianjin University, China
[2]Institute of linguistics, Chinese Academy of Social Sciences, China

`huyiying@tju.edu.cn, fenghui@tju.edu.cn, zqh2501387952@163.com, liaj@cass.org.cn`

## Abstract

Few studies have worked on the effects of tonal coarticulation and prosodic positions on the low rising tone in Xiamen Dialect. This study addressed such an issue. To do so, a new method, the Tonal Contour Analysis in Tonal Triangle, was proposed to measure the subtle curvature of the tonal contour. Findings are as follows: (1) The low rising tone in Xiamen Dialect has a tendency towards the falling-rising tone, which is significantly affected by the tonal coarticulation and prosodic positions. (2) The low rising tone presents as a falling-rising tone when preceded by a tone with a high offset, and as a low rising tone when preceded by a tone that ends up low. (3) The curvature of the low rising tone is greatest in the sentence-initial position, and is positively correlated to its own duration.

**Index Terms**: low rising tone, tonal coarticulation, prosodic position, Xiamen Dialect

## 1. Introduction

### 1.1. Effects of tonal coarticulation on tonal contours

Effects of tonal coarticulation are universal in tone languages such as Thai [1, 2], Vietnamese [3], Mandarin [4, 5, 6], Taiwanese [7, 8] and Malaysian Hokkien [9]. The influences of tonal coarticulation include carryover and anticipatory effects. Carryover effects occur when a tone is affected by the preceding tone, while anticipatory effects occur when a tone is affected by the following tone. Tonal interaction between neighboring tones can be assimilatory or dissimilatory. Previous studies on Mandarin [10], Taiwanese [11, 12, 13] and Vietnamese [3] found assimilatory tonal coarticulation that the fundamental frequency (F0) is higher when preceded or followed by a high tone than by a low tone. But the dissimilatory tonal coarticulation was also found in research on Thai [14], Taiwanese [8], and Mandarin [4].

Three effects of tonal coarticulation on tonal contours have been widely addressed in recent studies. First, the F0 range of tonal contour is affected by adjacent tones. The F0 range of a tone is higher when preceded by a high-offset tone than by a low-offset tone [8, 10]. Second, tonal coarticulation exerts an impact on F0 onsets and offsets of tonal contours. For example, the F0 offset of a rising tone in Mandarin is lower when preceding a high target tone than a low target tone [4]. Third, the F0 deviation can sometimes be influenced by neighboring tones in running speech. A prior study [7] classified tone contexts into conflicting contexts and compatible ones. The deviation of a tonal contour is relatively greater in the conflicting context where adjacent tonal values disagree than in the compatible context where adjacent tonal values agree [7].

### 1.2. Effects of prosodic positions on tonal contours

When it comes to effects exerted by prosodic positions, final-lowering and final-lengthening effects are widely explored. Evidence for final-lowering can be found in both tone languages and non-tone languages, including Mandarin [4], Taiwanese [8], Yoruba [15, 16], English [17, 18], Spanish [19], and Japanese [20, 21]. Previous studies indicated that F0 of the tone in utterance-final position was lower than in utterance-medial and utterance-initial positions, and the F0 is higher in the phrase-initial position [22-26]. In terms of final-lengthening effects which is considered a universal phenomenon, the duration of the tone is longer in utterance-final and phrase-final positions, shown in studies on Southern Bantu [27], Mandarin [28], and American English [29].

### 1.3. Low rising tones in Chinese dialects

Chinese dialects possess rich tonal systems, in which low rising tones perform different directions of diachronic changes. Low rising tones changed to falling-rising tones in Ningbo [30], Jinan [31, 32], Suzhou dialects [33] and Taiwanese [13], to middle rising tones in Shanghai dialect [34, 35], to high rising tones in Kaifeng dialect [36], to low level tones in Nanjing dialect [37], and to low falling tones in Jinan dialect [38, 39] and Yantai dialect [40].

The purpose of this study is to investigate the effects of tonal coarticulation and prosodic positions on tonal contours of low rising tones. Xiamen Dialect serves as a good subject for this purpose, as diachronic changes of low rising tone in Xiamen Dialect have been rarely explored while prior studies on Taiwanese have discovered the change from low rising tones to falling-rising tones among the young generation [13]. Given great phonological similarities between Taiwanese and Xiamen Dialect but discrepancies in their contact with Mandarin, the current study proposes a hypothesis: The low rising tone in Xiamen Dialect presents a tendency towards the falling-rising tone, and this unstable tendency is affected by tonal coarticulation and prosodic positions.

### 1.4. Tonal system of Xiamen Dialect

The tonal system of Xiamen Dialect comprises five smooth tones (T1, T2, T3, T4 and T5) and two checked tones (T6 and T7), listed in Table 1. All these seven tones have tonal alternations between base tones and sandhi tones. The tone group conditioning tone sandhi in Xiamen Dialect is delimited by the non-adjunct maximal projection [41, 42, 43]. Within a tone group, base tones are only produced at the right edge of the tone group while sandhi tones are produced in other positions [41].

Table 1: *Tonal system of Xiamen Dialect* [44].

|              | T1 | T2 | T3 | T4 | T5 | T6  | T7  |
|--------------|----|----|----|----|----|-----|-----|
| Base tones   | H  | LM | HL | L  | M  | M   | H   |
| Sandhi tones | M  | M  | H  | HL | L  | H/HL| L/L |

## 2. Methods

### 2.1. Participants

Eight speakers (4 female, 4 male) participated in the experiment, whose mean age was 20.5 (SD=0.5). The present study focuses on the young generation since they are the leader of language change and previous research indicated the low rising tone in Taiwanese changed to be falling-rising among younger speakers [13]. In recruiting, requirements were met to guarantee the typicalness and representativeness of participants. That is, all participants are bilingual speakers of Mandarin-Xiamen Dialect, born and raised in Xiamen, never left Xiamen for more than two months, and frequently use Xiamen Dialect in their daily life.

### 2.2. Materials

The target tone in this study, the low rising tone in Xiamen Dialect, is denoted as T2 in Table 1. The disyllabic target words listed in Table 2 include all possible tonal combinations of the low rising tone and non-checked tones in Xiamen Dialect, constructing five types of tonal coarticulation. These ten disyllabic target words were read in three carrier sentences "[Target word] is the word I want to read. I see [Target word] frequently. This word is [Target word]." Carrier sentences provide three types of prosodic positions, including sentence-initial, sentence-medial and sentence-final. All materials were read two times by eight participants, producing 480 tokens (10×3×2×8). In addition, the target tone in all types of tonal coarticulation and prosodic positions is produced as its base tone, as it is at the right edge of the tone group.

Table 2: *Disyllabic target words.*

| AO (σ1/σ2) | T2 | | |
|---|---|---|---|
| T1 | 开门 | /kʰui mŋ/ | 'open the door' |
|    | 中华 | /tiɔŋ hua/ | 'China' |
| T2 | 明年 | /mẽ nĩ/ | 'the next year' |
|    | 葡萄 | /pʰu tʰau/ | 'the grape' |
| T3 | 语言 | /gu giɛn/ | 'the language' |
|    | 水泥 | /tsui nĩ/ | 'the cement' |
| T4 | 臭虫 | /tsʰau tʰaŋ/ | 'the bug' |
|    | 菜头 | /tsʰai tʰau/ | 'the vegetable' |
| T5 | 问题 | /mbun tue/ | 'the problem' |
|    | 大门 | /tua mŋ/ | 'the gate' |

### 2.3. Procedure

Each participant was instructed to read materials presented on an iPad screen at fluent and normal speed in a quiet room. A Lenovo YOGA C740 computer and a Plantronics Blackwire 3220 Series headset microphone were used for the recording. The productions of speakers were recorded with PRAAT at 44.1 kHz 16 bits and a mono soundtrack [45].

### 2.4. Data analysis

#### 2.4.1. F0 analysis

For each target tone, the duration and F0 values of ten points were extracted by Voice Sauce [46]. The F0 values were normalized by Equation (1) which converts hertz to semitones [47, 48]:

$$F0^{\text{ST-Avg}} = \frac{12}{\log_{10} 2} \times \log_{10} \frac{xi}{ref} \qquad (1)$$

where *xi* refers to raw values of F0 as *i* takes the value from 1 to 10 for the ten measuring points, and *ref* is the mean pitch of each speaker. The resulting $F0^{\text{ST-Avg}}$ value is the semitone relative to the mean pitch of each speaker.

#### 2.4.2. Tonal Contour Analysis in Tonal Triangle

Since change in tonal curvatures serves as a clue of language change, to measure the subtle curvature of a tonal contour is necessary. And traditional acoustic analyses fail to measure subtle curvatures of tonal contours in the current study. On the one hand, tonal measurement based on the five-point number scale mainly serves for noticeable curvatures but not subtle ones [49]. On the other, MOTTA, although it accurately describes tonal values and contours [13], can not be applied to the study with larger samples. Thus, Tone Contour Analysis in Tonal Triangle (TCATT) is proposed for this study to measure the subtle curvature of a tonal contour.

The basic notion behind TCATT is that a tonal contour of a syllable can be treated as a triangle, which acoustically illustrates the tone shape of each tonal category [13]. TCATT comprises two steps. The first step is to classify the number of turning point (one or two in the present study) of a tonal contour, shown in Figure 1. If k ≥ $k_{min}$ and $k_{max}$, or k ≤ $k_{min}$ and $k_{max}$, the tonal contour has only one turning points. Otherwise, it has two or more turning points.

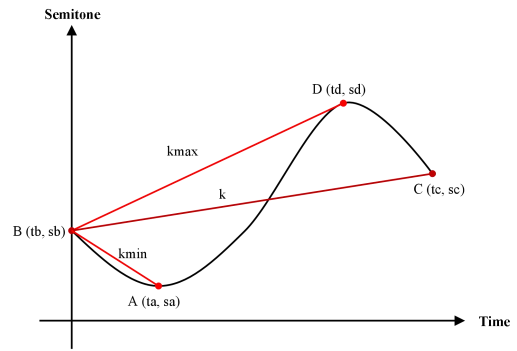

A: the lowest point of the tonal contour
B: the starting point of the tonal contour
C: the ending point of the tonal contour
D: the highest point of the tonal contour
k/ $k_{min}$/ $k_{max}$: the slope of the line connecting the starting point and the ending/ lowest/ highest point

Figure 1: *Step 1 of TCATT: Classifying the number of turning points of a tonal contour.*

The second step is to measure the curvature of the tonal contour. The turning point of a tonal contour constructs the tonal triangle, and tonal contours with one and two turning points are demonstrated respectively in Figure 2. Take the one-turning-point tone as an example (the left panel in Figure 2), the value of sine indicates the curvature of the tonal contour. The notion behind this is that the sine function is in a monotonically decreasing interval when values of cosine are negative in a tonal triangle. If cos∠A is negative, then ∠A is an obtuse angle. The larger the value of sin∠A is, the smaller the angle of ∠A is, the greater the curvature of this tonal contour is.

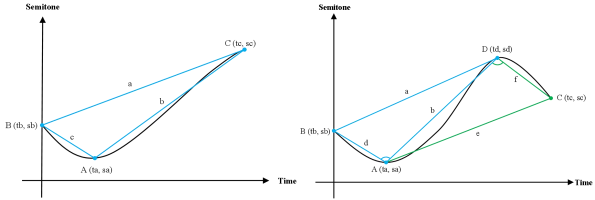

A: the first turning point of the tonal contour

B: the starting point of the tonal contour

C: the ending point of the tonal contour

D: the second turning of the tonal contour

$t_x$: the value of normalized time point of X (X= A, B, C, D)

$s_x$: the value of semitone of X (X= A, B, C, D)

y: the Euclidean distance between two points (y= a, b, c, d, e, f)

Figure 2: *Step 2 of TCATT: Measureing the curvature of the tonal contour.*

And The Law of Cosine (2) and the extension of The Law of Sine (3) are employed to calculate values of sine and cosine, taking a one-turning-point tone for instance.

$$\cos\angle A = \frac{b^2 + c^2 - a^2}{2bc} \quad (2)$$

$$\sin\angle A = \frac{\sqrt{2bc(1 - \cos\angle A)}}{c} \quad (3)$$

where ∠A refers to the angle A and a, b, c refer to the Euclidean distances between a, b, and c in the left panel of the tonal triangle in Figure 2.

## 3. Results and Discussion

The three-way ANOVA was conducted with the result of TCATT as the dependent variable, and tonal combinations, prosodic positions, and gender as independent variables. In terms of the number of turning points of target tones, only the effect of the gender is significant (F=37.212, *p*=0.000**<0.01). Male speakers produce more two-turning-point tones than females. And both female and male speakers tend to produce target tones with one turning point, rather than two. In terms of the curvature of target tones, the effect of the tonal coarticulation is significant (F=4.506, *p*=0.001**<0.01), so do the effect of the prosodic position (F=13.619, *p*=0.000**<0.01) and the effect of gender (F=77.400, *p*=0.000**<0.01). Figure 3 shows normalized tonal contours of target low rising tones with different independent variables.

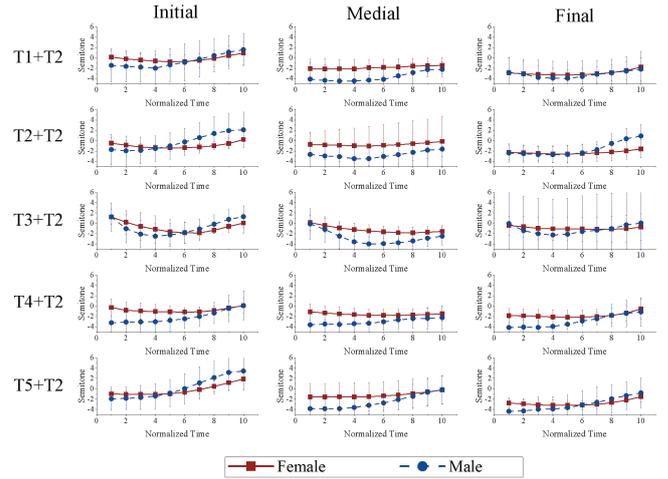

Figure 3: *Tonal contours of target low rising tones with different tonal combinations, prosodic positions and gender groups.*

### 3.1. Effects of tonal coarticulation on tonal contours

The effect of tonal coarticulation on tonal contours of low rising tones is significant (F=4.506, *p*=0.001**<0.01). The curvature of the target low rising tone in different tonal combinations is illustrated in Figure 4, with values of sine indicating the curvature of tonal contours. The curvature of the target low rising tone is greatest in the T3T2 combination.

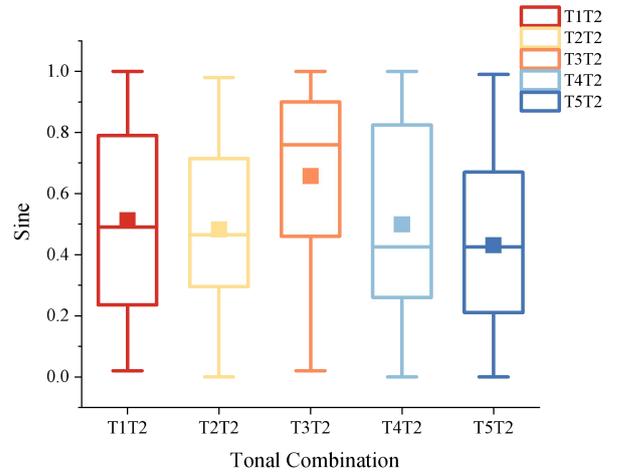

Figure 4: *Values of sine with different tonal combinations.*

To explore carryover effects, tonal contours of disyllabic target words of five tonal combinations are demonstrated in Figure 5 (upper panel for female speakers, and lower panel for male speakers). The second syllable is the target low rising tone, to be combined with varied first syllables of the five smooth tones in Xiamen Dialect. And the prosodic position is controlled to be sentence-medial.

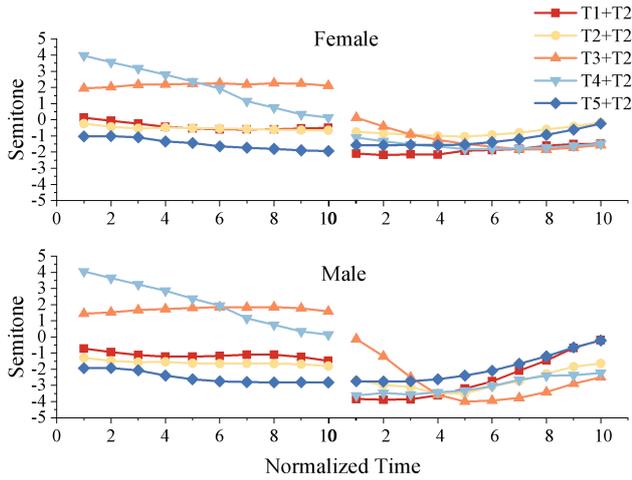

Figure 5: *Tonal contours of disyllabic target words in different tonal combinations.*

Carryover effects occur when the target low rising tone is preceded by the high level tone T3$_{sandhi}$, as the higher offset of the first syllable led to the higher onset of the second syllable, shown in orange lines. The low rising tone in Xiamen Dialect performs as falling-rising if the preceding syllable has a high offset and performs as low-rising if the preceding syllable ends up low. Such findings reach a consensus with the prior study on Taiwanese [13]. The present study also suggests that the curvature of the target low rising tone produced by males is greater than females ($z$=-7.944, $p$=0.000**<0.01), especially when the low rising tone is preceded by a high level tone, shown in orange lines in Figure 5.

### 3.2. Effects of prosodic positions on tonal contours

The effects of prosodic position on tonal contours of the low rising tone is significant (F=13.619, $p$=0.000**<0.01). Figure 6 shows the curvature of the target low rising tone in different prosodic positions, suggesting the greatest curvature in the sentence-initial position.

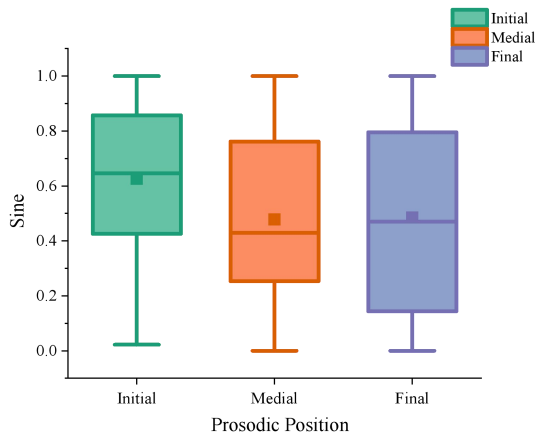

Figure 6: *Values of sine with different prosodic positions.*

To explain effects exerted by prosodic position, the duration of the target low rising tone in different prosodic positions is illustrated in Figure 7.

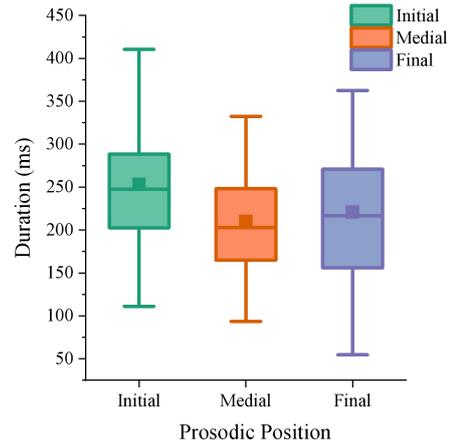

Figure 7: *Duration of target low rising tones with different prosodic positions.*

The results indicate that both duration and the value of sine in the sentence-initial position are greater than those in the sentence-final position, which is greater than those in the sentence-medial position. Based on this, a hypothesis is proposed that the curvature of the low rising tone is positively correlated to its duration, which is verified by a Pearson correlation coefficient of 0.522 ($p$=0.000**).

## 4. Conclusions

The purpose of the study is to explore the effects of the tonal coarticulation and prosodic positions on tonal contours of the low rising tone in Xiamen Dialect.

To measure the tonal contour whose curvature is too subtle to be measured by traditional methods, the study proposes Tonal Contour Analysis in Tonal Triangle (TCATT). With our proposed method, the study finds that the low rising tone in Xiamen Dialect has a tendency towards the falling-rising tone, which is significantly affected by the effects of tonal coarticulation and prosodic position. In terms of tonal coarticulation, the low rising tone performs as a falling-rising tone when it is preceded by a tone with a high offset, and it presents as a low rising tone when the preceding tone ends up low. In terms of prosodic positions, the curvature of the low rising tone is the greatest in the sentence-initial position. A positive correlation between the curvature of the low rising tone and its duration is also suggested by the present study.

In addition, there is an interesting observation in the present study that the curvature of the low rising tone produced by males is greater than those by females, and male speakers sometimes produce the low rising tone with two turning points. Subtle as these curvatures are, young male speakers in Xiamen Dialect can be supposed as the leader of the potential on-going change from the low rising tone to the falling-rising tone, as the same change has been discovered in Taiwanese [13]. Future research is to focus on this potential language change in Xiamen Dialect and to investigate the social factors behind it.

## 5. Acknowledgements

This work is supported partly by the "Four Batches" Talent Project "Intonation typology" and partly by the National Key R&D Program of China under Grant 2018YFB1305200.